\documentclass[12pt]{article}
\usepackage{graphicx}
\usepackage{pazh}

\usepackage{lscape}
\usepackage{amssymb}
\usepackage[T2A]{fontenc}
\usepackage[russian]{babel}

\tightenlines

\def\arcsec{$^{\prime\prime\,}$}
\def\arcmin{$^{\prime\,}$}

\def\Ю{$^{\mbox{\small Ю}}$}
\def\А{$^{\mbox{\small А}}$}
\def\Б{$^{\mbox{\small Б}}$}
\def\Ц{$^{\mbox{\small Ц}}$}

\begin{document}

\centerline{\it Will be published: Astronomy Letters, 2012, Vol. 38, No. 10, pp. 629-637 }

\title{\bf Accurate Localization and Identification of Six Hard X-ray Sources from \emph{Chandra} and \emph{XMM-Newton} data}

\author{\bf \hspace{-1.3cm}\copyright\,2012 \ \
D. I. Karasev\affilmark{1}$^{\,*}$, A. A. Lutovinov\affilmark{1}, M. G. Revnivtsev\affilmark{1} and  R. A. Krivonos\affilmark{2,1}}
\affil{
$^1$ {\it Space Research Institute, ul. Profsoyuznaya 84/32, Moscow, 117997 Russia} \\
$^2$ {\it Max Planck Institut fur Astrophysik, Karl-Schwarzschild-Str. 1, Postfach 1317, D-85741 Garching, Germany} \\ }
\vspace{2mm}

\sloppy
\vspace{2mm}
\noindent

\sloppy

            We present the results of Chandra and XMM-Newton observations for six hard X-ray sources
(IGR\,J12134-6015, IGR\,J18293-1213, IGR\,J18219-1347, IGR\,J17350-2045, IGR\,J18048-1455, XTE\,J1901+014)
 from the INTEGRAL all-sky survey. Based on these observations, we have improved
significantly the localization accuracy of the objects and, therefore, have managed to identify their optical
counterparts. Using data from the publicly available 2MASS and UKIDSS infrared sky surveys as
well as data from the SOFI/NTT telescope (European Southern Observatory), we have determined the
magnitudes of the optical counterparts, estimated their types and (in some cases) the distances to the
program objects. A triplet of iron lines with energies of 6.4, 6.7, and 6.9 keV has been detected in the X-ray
spectrum of IGR J18048-1455; together with the detection of pulsations with a period of $\sim$1440 s from
this source, this has allowed it to be classified as a cataclysmic variable, most likely an intermediate polar.
In addition, broadband X-ray spectra of IGR\, J12134-6015 and IGR J17350-2045 in combination with
infrared and radio observations suggest an extragalactic nature of these objects. The source IGR J18219-1347 presumably belongs to the class of high-mass X-ray binaries.

\bigskip


\vfill

{$^{*}$ E-mail: dkarasev@iki.rssi.ru}

\newpage

\section*{INTRODUCTION}
The value of astronomical sky surveys depends significantly on how complete (according to various criteria) they are. In particular, the INTEGRAL all sky hard X-ray (17-60 keV) survey is currently the most complete in its coverage of the Galactic plane (see, e.g., Krivonos et al. 2010a, 2012). It should be noted that there are much deeper (more sensitive) sky surveys in this region. However, they cover a small area, which limits the application of their results for various types of astrophysical problems. The completeness of determining the nature of objects is one of the most important properties of sky surveys. In this paper, we continue our studies to determine the nature of hard X-ray sources detected in the INTEGRAL survey. Our sample includes the following objects: IGR J12134-6015, IGR J18293-1213, IGR J18219-1347, IGR J17350-2045, IGR J18048-1455, and XTE J1901+014 from the currently available catalog of the Galactic plane (Krivonos et al. 2012). All these objects have been studied poorly and their nature has not been established.

\section*{OBSERVATIONS}

Our study is based on data from the \emph{Chandra} (ObsID. 12498, 12499, 12500, 12501) and \emph{XMM-Newton} (ObsID. 06554901, 04053903) X-ray observatories. Thanks to the \emph{HRC/Chandra} telescope, four X-ray objects were localized. The X-ray spectra obtained from \emph{XMM-Newton} data for two more sources were studied to establish their nature using the standard SAS v11.0\footnote{http://xmm2.esac.esa.int/sas/} software. To construct broadband spectra of the objects, we also used data from the INTEGRAL observatory (Winkler et al. 2003, the \emph{IBIS} telescope), processed with the specially developed software (Krivonos et al. 2010b), and \emph{Swift} one (Gehrels et al. 2004, the \emph{XRT} telescope); software from the HEASOFT\footnote{http://heasarc.gsfc.nasa.gov/docs/software/lheasoft/} package was used for their subsequent analysis. For optical identifications, we used data from the 2MASS\footnote{http://www.ipac.caltech.edu/2mass/} and \emph{UKIDSS}\footnote{http://www.ukidss.org/index.html} infrared sky surveys as well as observations with the \emph{SOFI} instrument of the NTT telescope\footnote{http://archive.eso.org}, which are publicly available.

We reprocessed all these data, except for the 2MASS catalog, using the method of PSF photometry and the DAOPHOT III software from the SCISOFT/ESO package \emph{SCISOFT/ESO}\footnote{http://www.eso.org/sci/software/scisoft/}. The photometric solutions were obtained using 2MASS as a reference catalog. For this purpose, we selected 500 bright but not overexposed stars from our data and then identified them with stars from the 2MASS catalog using subroutines from the WCStools 3.8.4 package. In this way, we determined the photometric constants for the J, H, Ks bands.
The PSF photometry (DAOPHOT) parameters were chosen so as to minimize the photometric errors.

\bigskip
\hspace{-7mm}\begin{table}[]

\footnotesize{
   Table 1. The list of hard X-ray sources localized from Chandra and XMM-Newton data
   \medskip
   ~\hspace{-17mm}\begin{tabular}{lccccc}
     \hline
     \hline
          Name &  R.A. & Dec & &Magnitudes of the counterparts   \\
& (J2000) & (J2000) & J & H & Ks\\[1mm]
     \hline
     IGR\,J12134-6015$^{Ch}$ & 12$^h$13$^m$23.98$^s$ &  -60\deg15\arcmin16.6\arcsec & $16.45\pm0.12$ & $>16.5$ & $>15.5$ \\
     IGR\,J17350-2045$^{XMM}$ & 17$^h$34$^m$58.95$^s$ &  -20\deg45\arcmin31.5\arcsec & - &  $14.92\pm0.06$ & $14.37\pm0.06$ \\
     IGR\,J18048-1455$^{XMM}$ & 18$^h$04$^m$38.80$^s$ &  -14\deg56\arcmin46.6\arcsec & $16.00\pm0.10$ & $15.27\pm0.12$ & $>14.3$\\
     IGR\,J18219-1347$^{Ch}$ & 18$^h$21$^m$54.82$^s$ &  -13\deg47\arcmin26.7\arcsec & - & - & -\\
     IGR\,J18293-1213$^{Ch}$ & 18$^h$29$^m$20.16$^s$ &  -12\deg12\arcmin50.7\arcsec & $16.81\pm0.03$ & $15.97\pm0.04$ & $15.53\pm0.04$\\
     XTE\,J1901+014$^{Ch}$ & 19$^h$01$^m$40.20$^s$ &  +01\deg26\arcmin26.5\arcsec & $>21$ & $>20.5$ & $>19$ \\

     \hline
    \end{tabular}
    \medskip
    $^{Ch}$ -- the localization accuracy is 0.64\arcsec\ (\emph{Chandra})\\
    $^{XMM}$ -- the localization accuracy is 2\arcsec\ (\emph{XMM-Newton})
    }
\end{table}

\section*{LOCALIZATION}

As a response to the request of our group (no. 12400892), the \emph{Chandra} observatory (HRC instrument) performed observations of several X-ray objects, candidates for high-mass X-ray binaries, in 2011. The list of targets also included XTE J1901+014, whose nature is not completely clear and an additional improvement of the source's astrometric position could clarify the situation. The localization accuracy of all these sources was determined by the accuracy with which the telescope's orientation was known, because we had no other bright objects in the fields of view under study based on which an additional correction of the astrometric referencing could be made. Thus, we assume that the localization accuracy of the sources is 0.64 (see the table). Such an accuracy allowed us to determine their optical counterparts using \emph{2MASS} and \emph{UKIDSS} data. In addition, we also analyzed two objects (IGR J17350-2045 and IGR J18048-1455) that were localized with an accuracy of  $\simeq2$\arcsec\ using \emph{XMM-Newton} data and for which the optical identification was also made unambiguously.

\newpage

\section*{RESULTS}

\subsection*{XTE\,J1901+014}

XTE\, J1901+014 was discovered as a result of itsbright outburst in 2002 (Remillard and Smith 2002). A similar burst from the source was also detected in 2010 by the Swift observatory (Krimm et al. 2010). Outside its outburst activity, the source is stably detected by the \emph{IBIS} telescope of the \emph{INTEGRAL} observatory in the 17-60 keV energy band with a flux of about 2.3 mCrab (Karasev et al. 2007). The nature of the object is still not completely clear; different authors agree that XTE J1901+014 is most likely a low-mass X-ray binary (Smith et al. 2007; Karasev et al. 2008; Torrejon et al. 2010).
Using data from the Chandra observatory (the \emph{HRC} instrument), we localized the object with coordinates R.A.=19$^h$01$^m$40.20$^s$ , Dec = Dec=+01\deg26\arcmin26.5\arcsec\ (J2000) and an accuracy of $\simeq$0.64\arcsec (Fig. 1), in good agreement with the previous measurements based on \emph{XMM-Newton} data (Smith et al. 2007; Karasev et al. 2008). Despite the high localization accuracy of the source, we nevertheless failed to determine its optical or infrared counterpart, because even the UKIDSS survey is not deep enough to obtain its statistically significant magnitude at least in one band (we see from Fig. 1 that the \emph{Chandra} error box of the source does not coincide with any of the three presumed optical objects that fell into the \emph{XMM-Newton} error box).
At such upper limits on the brightness of the infrared counterpart, we may conclude that it cannot be a giant star; otherwise it would be brighter than the observed upper limits even if it were located at the opposite side of the Galaxy ($\simeq$22 kpc). This confirms the previously made assumptions that XTE J1901+014 is a low-mass X-ray binary (LMXB). If the system is actually an X-ray binary accreting matter from a low-mass star through the inner Lagrangian point, then we can estimate its possible distance and orbital period using some additional information about such binaries. The optical luminosity of the disk in LMXBs (accreting matter through the inner Lagrangian point) is considerably higher than that of the companion star. Studies of LMXBs have revealed a certain relation between their X-ray luminosity, orbital period, and optical (infrared) luminosity. It stems from the fact that the optical (infrared) luminosity in LMXBs results from the reprocessing of X-rays from the compact object in the accretion disk, whose size depends on the orbital period of the binary (van Paradijs and McClintock 1994; Revnivtsev et al. 2012):

\begin{figure}
\begin{centering}
\includegraphics[width=9cm,bb=30 150 580 645]{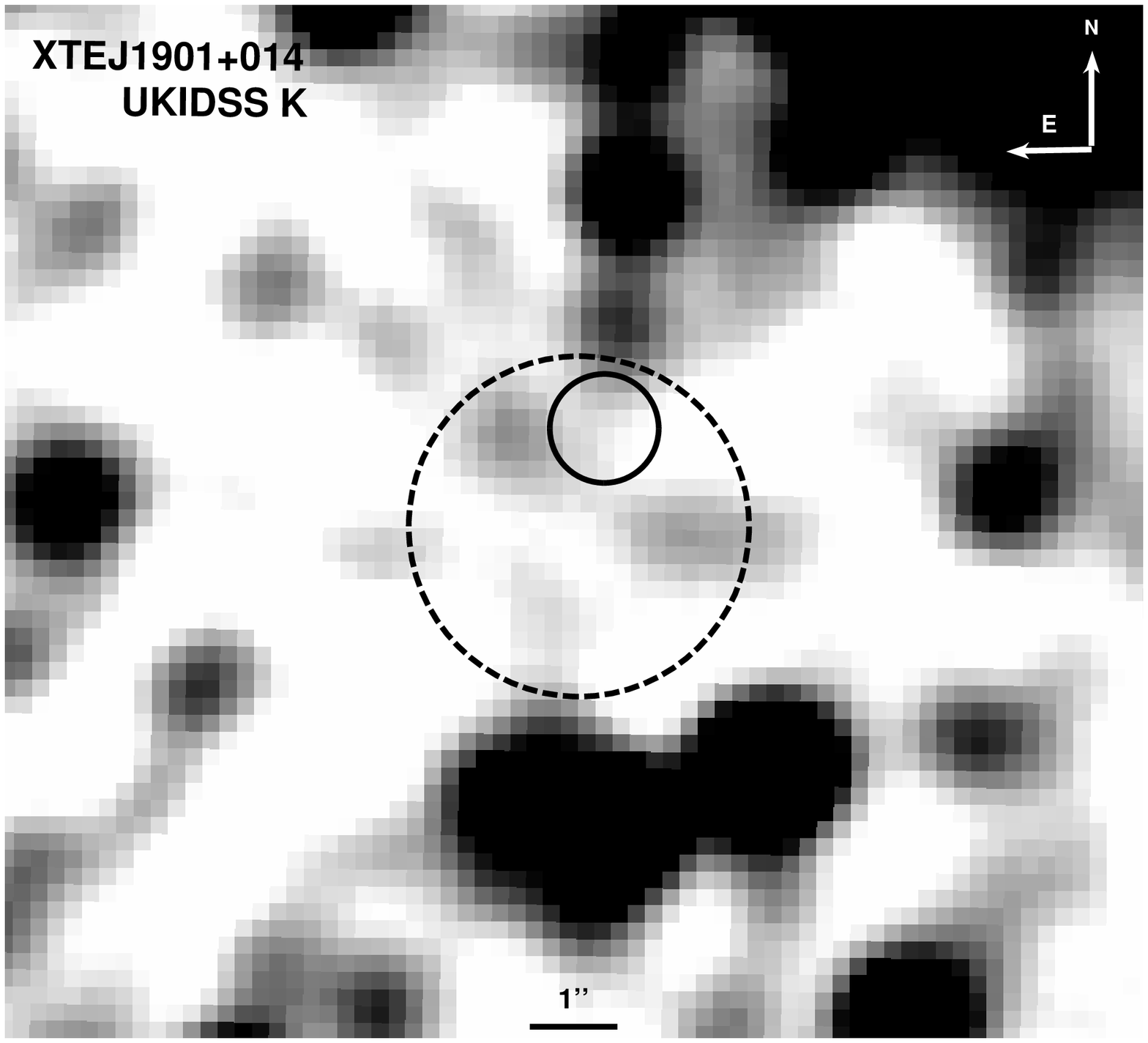}

Fig. 1. Image of the sky around the X-ray source XTE J1901+014 obtained from \emph{UKIDSS} data in the K band. The large and small circles indicate the uncertainty in the object's position from \emph{XMM-Newton} (the radius is 2\arcsec,, Karasev et al. 2008) and \emph{Chandra} (the radius is 0.64\arcsec, this paper) data, respectively.\label{xtej1901_ima}
\end{centering}
\end{figure}

$$
M_{K}=(2.66\pm0.11)-2.5\log \Sigma_{K}
$$

\noindent
where $\Sigma_{K}=(L_{\rm X}/L_{\rm Edd})^{0.29} P({\rm h})^{0.92}$, $M_{K}$ --  the system's absolute magnitude in the $K$ band, $L_{\rm X}$ -- its X-ray luminosity,  $L_{\rm Edd}$ -- the Eddington luminosity limit for a neutron star and $P({\rm h})$ -- the orbital period of the system in hours.

Suppose that the compact object in the binary system is a neutron star with a mass $M\simeq1.4 M_{Sun}$ and an Eddington luminosity limit $L_{Edd}\simeq1.5\times10^{38}$. The flux from the system in quiescence in the $0.6-12$ keV energy band is $\simeq2.5$ mCrab (Karasev et al. 2008). For the object's given distance and orbital period, we can obtain its infrared brightness (Fig. 2).

Since there is no giant star in the binary capable of producing an absorbing envelope around it, the column density of the matter derived by fitting its X-ray spectrum ($N_H\simeq 2.6\times10^{22}$ cm$^{-2}$ Karasev et al. 2007, 2008), to a first approximation, gives an estimate of the interstellar extinction toward the object $A_{Ks}=0.112A_V\sim0.112 \times N_HL/1.9\times10^{21}$ cm$^{-2}$   $\approx1.5$ (here, we used the relation $A_V\sim N_H/1.9\times10^{21}$ cm$^{-2}$ from Predehl and Schmidt (1995)). Thus, our upper limit on the object's observed brightness after its correction for the interstellar extinction is $Ks>17.5$. The dotted lines mark the domain of distances at which an extinction $A_{Ks}$ = 1.5 is accumulated toward the source (within a radius of 30' ) (Marshall et al. 2006). The thin vertical line (P $\sim$2 h) roughly corresponds to the separation of binaries with degenerate and nondegenerate companions. The thick polygon indicates the region bounding the binary's possible distances and periods under the assumption that the period-X-ray luminosity relation is defined by the law $1.8P^{2.5}\times10^{35}$erg s$^{-2}$ $<L_{X}<7P^{2.5}\times10^{37}$ erg s$^2$ (see Revnivtsev et al. 2011; Iben and Tutukov 1984). The typical orbital periods for low-mass binaries with a normal, nondegenerate companion star (not a giant) lie within the range from $\sim2$ to $\sim$10 h, while the binaries with shorter periods generally have a degenerate companion (this boundary is indicated in Fig.2 by the thin solid line). When taking into account all of the above constraints and assuming that the relativistic companion of XTE J1901+014 is a neutron star, the domain of admissible orbital periods and distances to the binary is indicated in Fig.2 by the gray color if the second companion in the binary is nondegenerate in nature. Number 1 in the figure marks the domain of values for the corresponding parameters under the assumption that the companion of the neutron star is a white dwarf. In conclusion, it should be noted that if the compact object in XTE J1901+014 is a black hole, then the constraints on the distance related to the attainment of the Eddington limit during the 2002 burst are removed, and the object itself can belong to the class of fast X-ray novae, like the sources V4641 Sgr and Swift J195509.6+261406 (for more detail, see Kasliwal et al. 2008).

\begin{figure}
\begin{centering}
\includegraphics[width=10cm,bb=100 200 470 600]{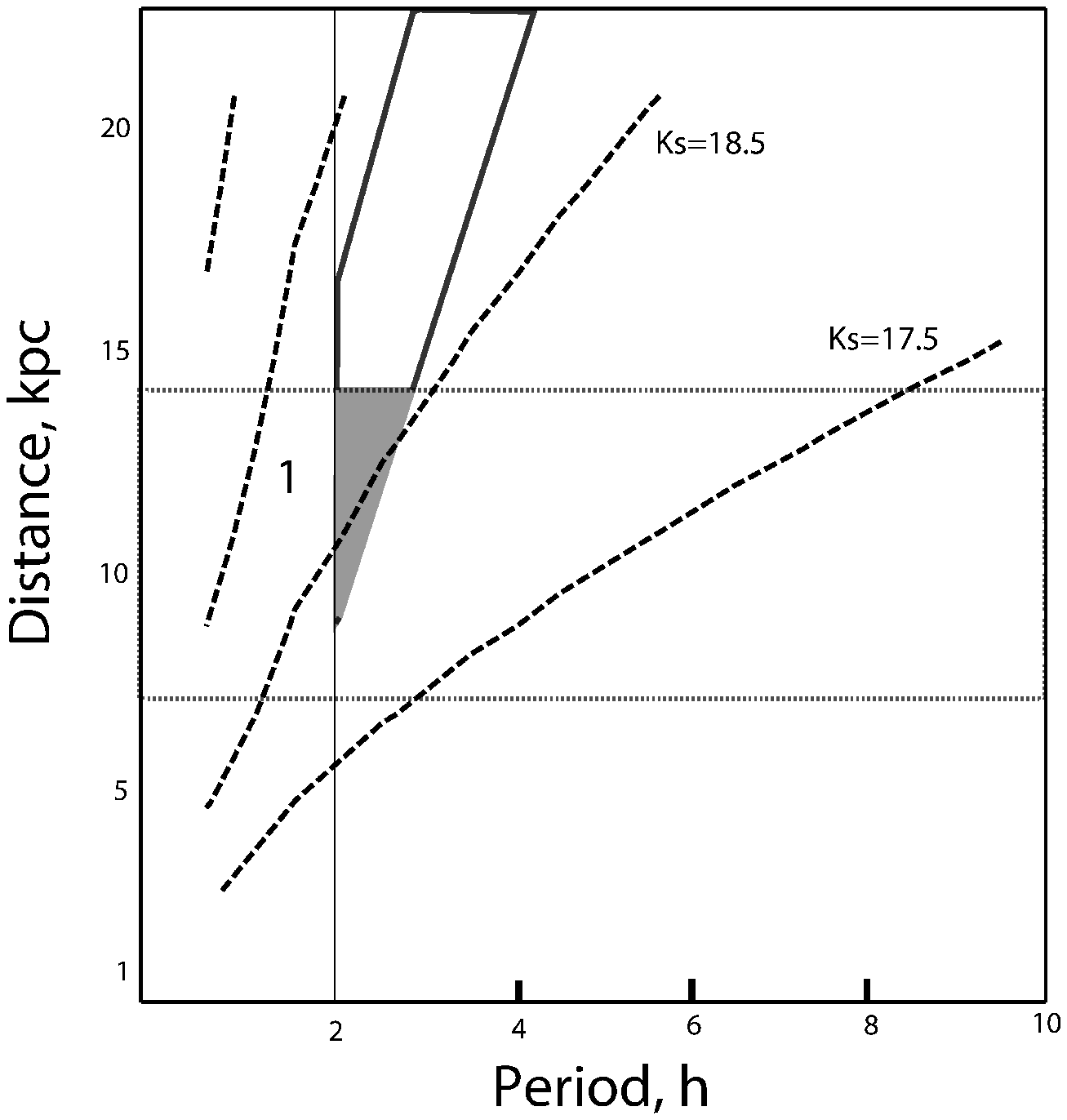}

Fig.2 Constraints on the distance and orbital period of the binary XTE J1901+014 under the assumption that it is an LMXB. The dashed lines indicate the Ks-band optical magnitudes of the system calculated for different sets of P and D. The dotted lines mark the domain of distances at which an extinction $A_{Ks}=1.5$ is accumulated toward the source (within a radius of 30\arcmin) (Marshall et al. 2006). The thin vertical line ($P\simeq2$ h) roughly corresponds to the separation of binaries with degenerate and nondegenerate companions. The thick polygon indicates the region bounding the binary's possible distances and periods (for more detail, see the text). The domain of possible P and D for XTE J1901+014 with a nondegenerate companion is highlighted by the gray color; number 1-if the companion is a white dwarf.\label{porb_dist}
\end{centering}
\end{figure}

\subsection*{IGR\,J12134-6015}

\begin{figure*}
\begin{centering}
\vbox{
\includegraphics[width=8cm,bb=30 150 580 645]{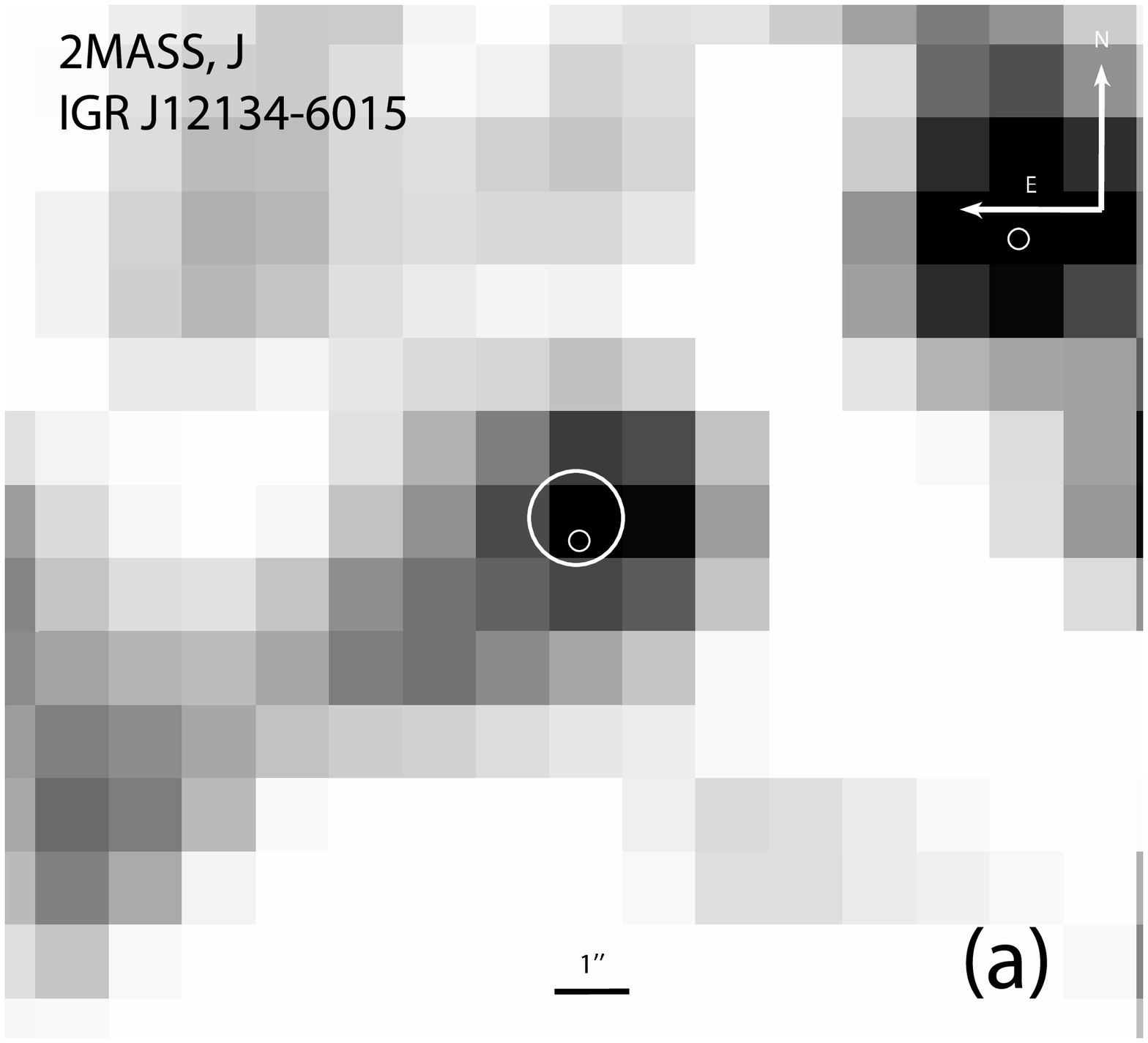}
\includegraphics[width=8cm,bb=30 150 580 645]{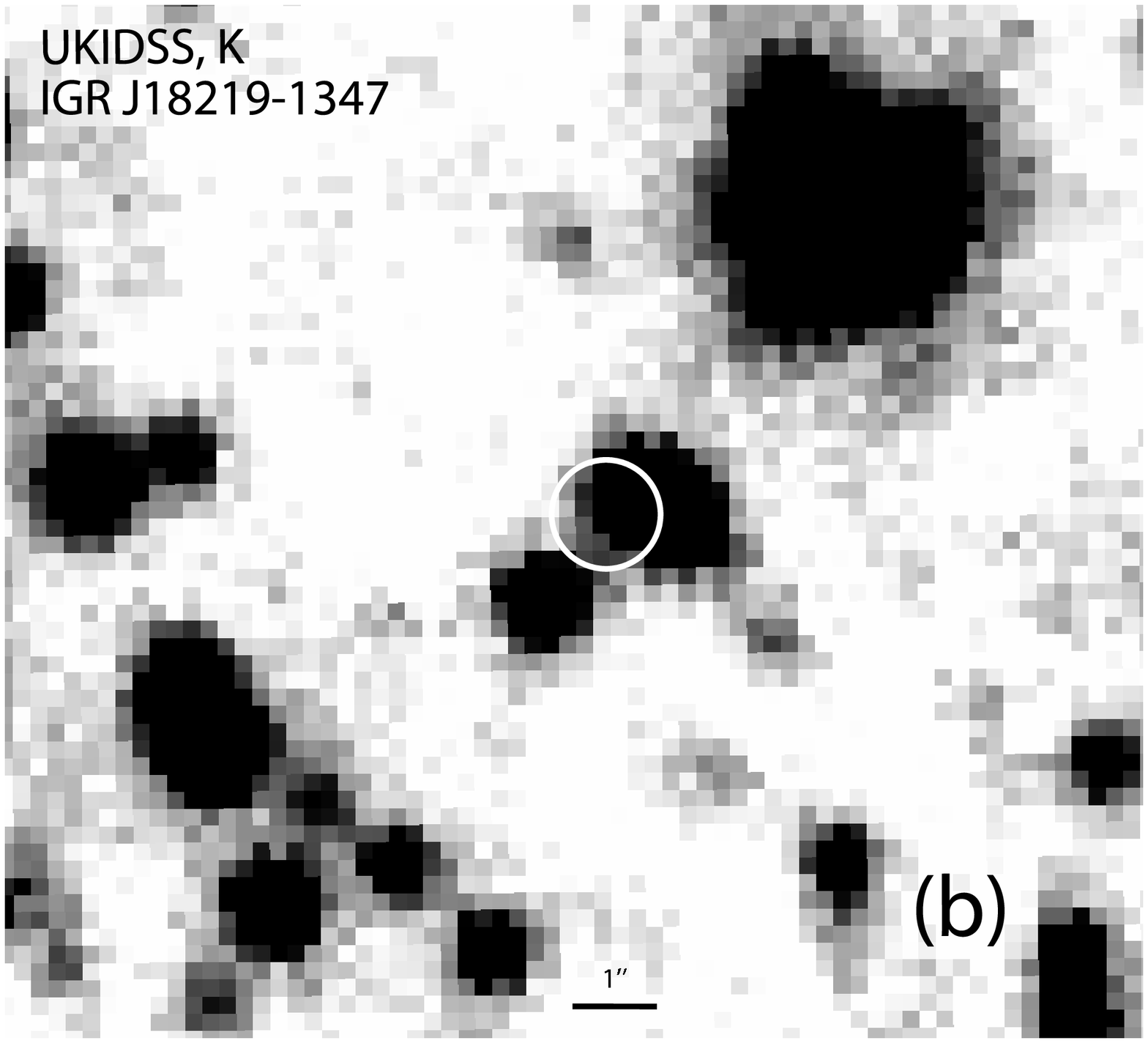}
\includegraphics[width=8cm,bb=30 150 580 645]{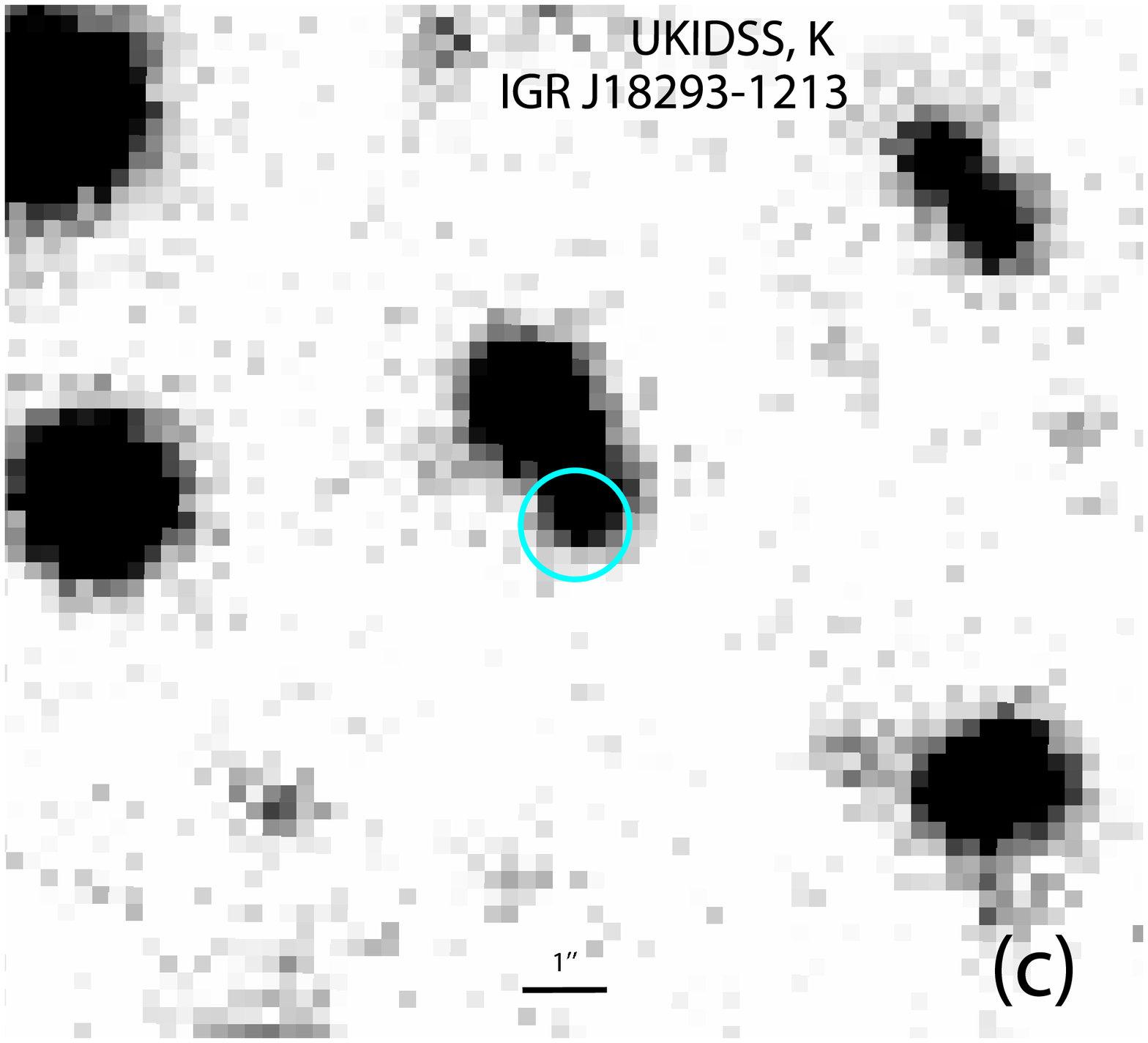}
\includegraphics[width=8cm,bb=30 150 580 645]{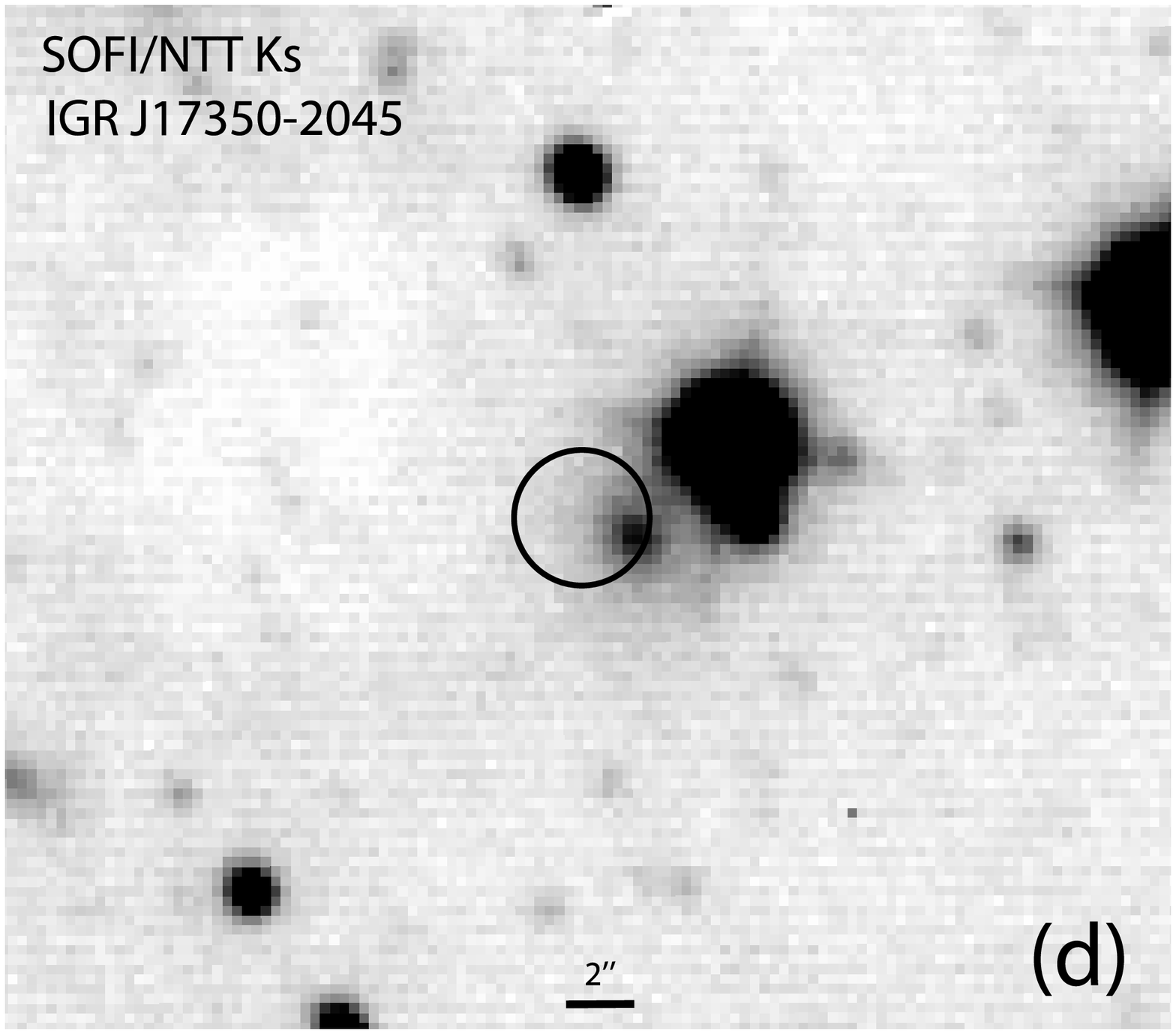}

}

Fig.3 Images of the sky regions around IGR J12134-6015 (a), IGR J18219-1347 (b), IGR J18293-1213 (c), and IGR J17350-2045 (d) obtained in the infrared wavelength ranges from \emph{UKIDSS} and \emph{SOFI/NTT} data. The large circles indicate the \emph{Chandra} (0.64\arcsec , a-c) and \emph{XMM-Newton} (2\arcsec, d) error boxes. The small circles in panel (a) indicate the positions of the infrared sources from the \emph{2MASS} catalog. \label{image_infra}
\end{centering}
\end{figure*}

In the error box of the X-ray source IGR J12134-6015, only one object with coordinates R.A.=12$^h$13$^m$23.98$^s$, Dec=-60\deg15\arcmin16.9\arcsec\ (J2000) (at a distance of 0.32\arcsec\ from the localization center of the X-ray source, Fig. 3a) is present in the 2MASS infrared sky survey. Its brightness is $J=16.45\pm0.12$; no counterpart is detected at a statistically significant level in the H and Ks bands (only the upper limits $H > 16.5$ and $Ks > 15.5$ are given in the 2MASS catalog). The corresponding upper limits for the colors are $J - H < 0.07$ and $J - Ks < 1.07$. If IGR J12134-6015 is assumed to be a binary with a massive companion, then it follows from the constraints on the color of the infrared object that the optical companion in the binary can be a star not later than the $A$ type (in the absence of interstellar extinction). The broadband spectrum of the source obtained with \emph{INTEGRAL} (17-100 keV) and \emph{SWIFT} (0.5-10 keV, ObsID.00038348) data shows no influence of interstellar extinction (the 2$\sigma$ upper limit is $\simeq4\times10^{21}$ atoms cm$^-2$ ) and can be described by a power law for the spectral density of photons in the form $dN(E)/dE\propto E^{-0.85}$ without an exponential cutoff up to energies of $\sim100$ keV (Fig. 4). Such a hard energy spectrum is quite atypical for galactic sources, but it is often encountered among extragalactic sources. Thus, the object being studied is most likely extragalactic in nature and is an active galactic nucleus or a blazar. Optical spectroscopic observations are needed for definitive conclusions to be reached.

\begin{figure*}
\begin{centering}
\includegraphics[width=10cm,bb=15 270 565 690]{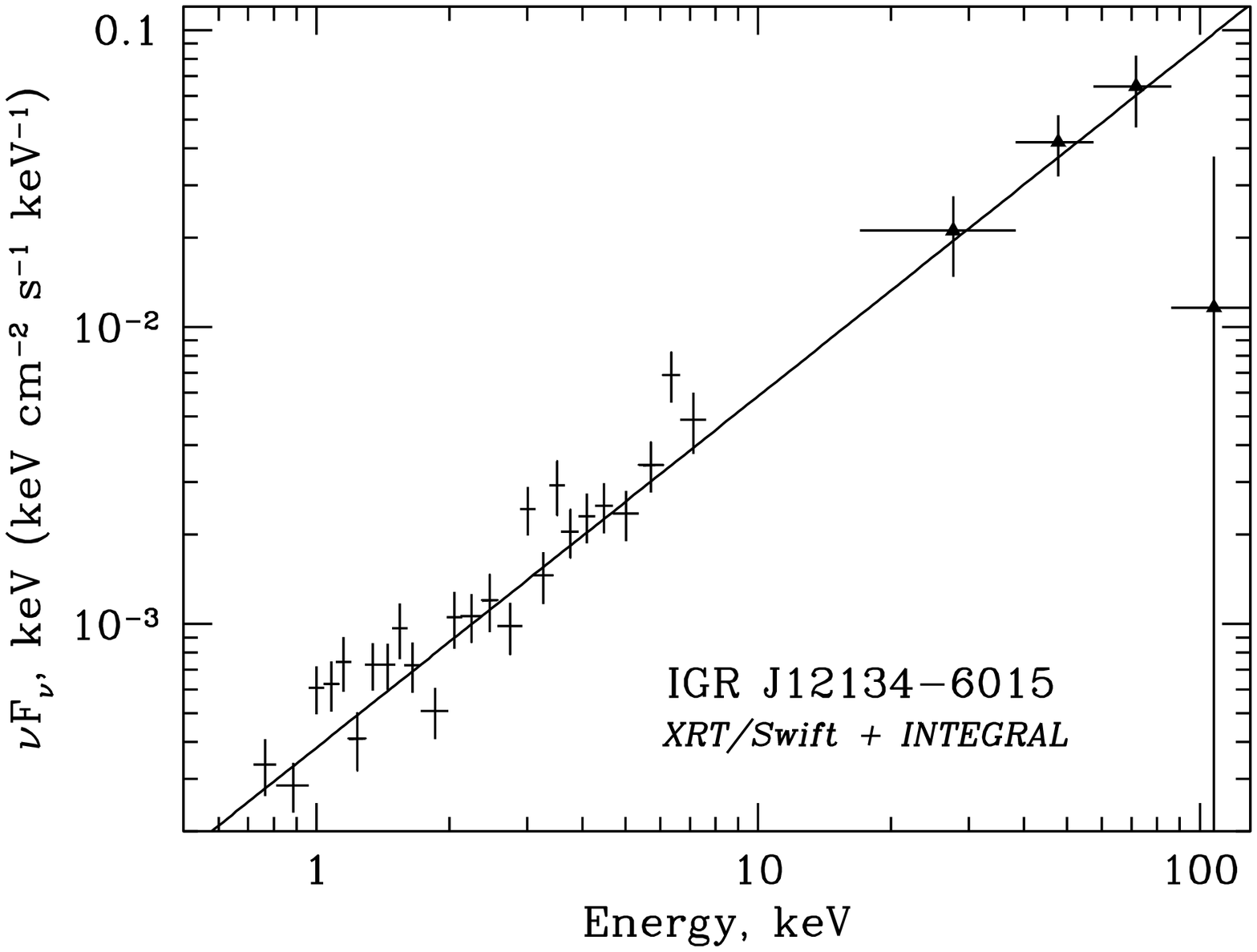}

Fig.4 X-ray energy spectrum of IGR J12134-6015 from \emph{XRT/Swift} (crosses) and \emph{IBIS/INTEGRAL} (triangles) data. The solid line is the best power-law fit with a slope $\Gamma\simeq 0.85$.\label{igrj12134_spectr}
\end{centering}
\end{figure*}

\subsection*{IGR\,J18293-1213}

In the \emph{Chandra} error box of the source (see the table), there is only one object in the \emph{UKIDSS} survey with coordinates R.A.=18$^h$29$^m$20.153$^s$, Dec=-12\deg12\arcmin50.44\arcsec\ (J2000) (0.26\arcsec\ from the X-ray localization center) and the following magnitudes in three infrared bands: $J=16.81\pm0.03$, $H=15.97\pm0.04$, $Ks=15.53\pm0.04$ (they were determined using PSF photometry of the original \emph{UKIDSS} data and \emph{2MASS} as a reference catalog). If the source is a high-mass binary whose optical and infrared brightness is determined by the companion star, then we can attempt to determine its spectral type by comparing the colors of our object and those of stars of known types. The gray dots in Fig. 5 indicate the colors of stars of different types (without any influence of interstellar extinction, the \emph{Hipparcos} catalog; van Leeuven and Fantino 2005). The strips at the bottom indicate the regions of the diagram corresponding to a particular type of stars (marked by the letters). The rectangle indicates the position of IGR J18293-1213 in the diagram according to its observed magnitudes and their errors. Correcting the source's magnitudes for different extinctions ($A_{Ks}$, see the labels in the figure), we shift the rectangle downward and to the left toward unreddened stars. The direction of motion of the rectangle is indicated by the dashed straight lines whose slope corresponds to the ratio $E(J-Ks)/E(H-Ks)$ under the assumption of a standard extinction law (Cardelli et al. 1989). Thus, only the stars lying within the dashed lines can be the source's companions. The corners indicate the shift in the source's position when changing the extinction by $\Delta A_{Ks}=0.25$ and $0.5$ ($\Delta A_{Ks}=0$ ($\Delta A_{Ks}=0$ corresponds to the observed position of the source); in this case, $E(J-K)=1.46\times A_K$, $E(J-H)=0.88\times A_K$. Note that according to Marshall et al. (2006), the extinction $A_{Ks}=1$ toward the source is accumulated at a distance of $\simeq4$ kpc.

\begin{figure}
\begin{centering}
\includegraphics[width=10cm,bb=60 200 570 700]{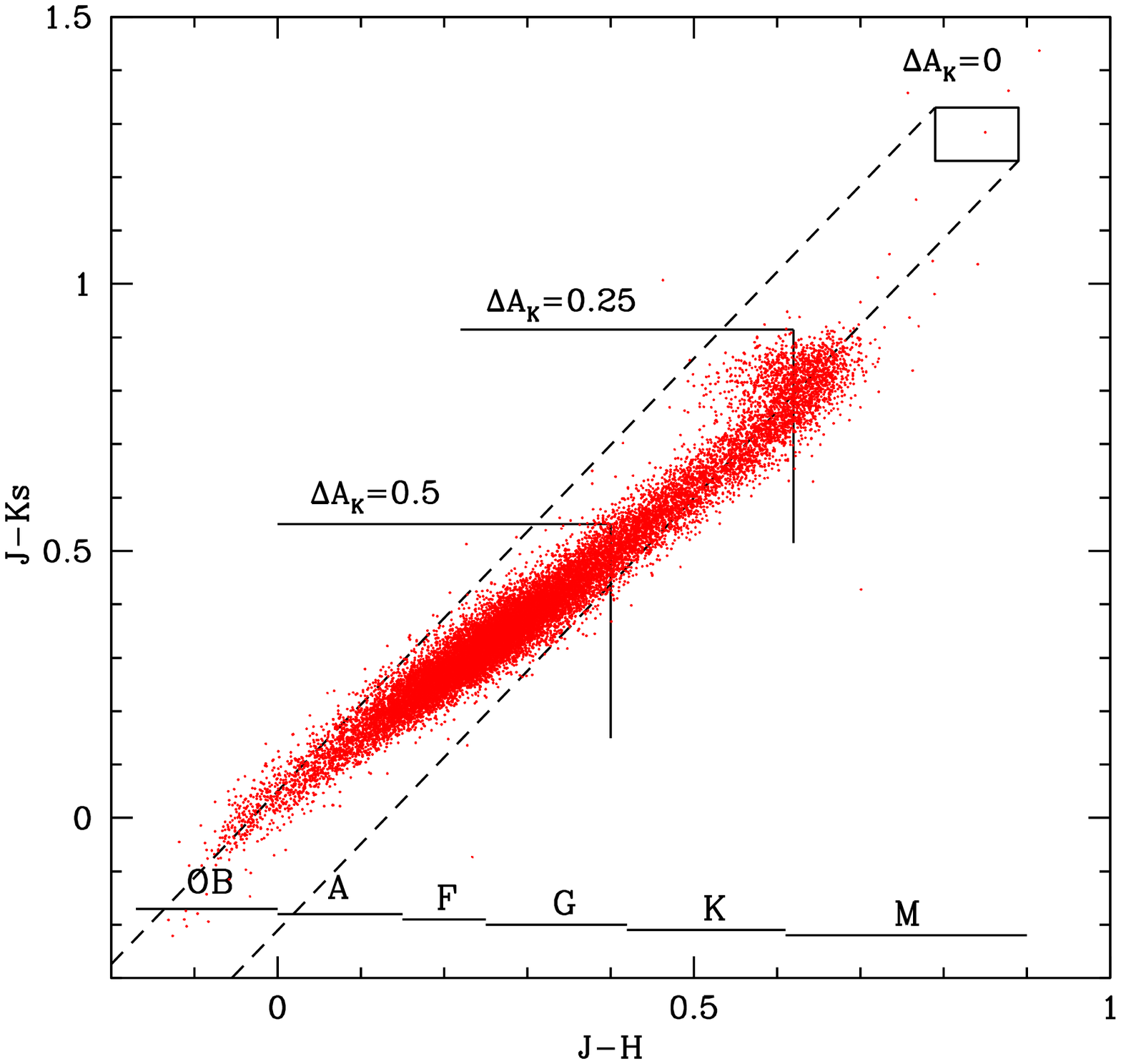}

Fig.5 Color-color diagram for unreddened stars of various spectral types (marked by the strips with the letters of spectral types). The rectangle indicates the domain of colors for IGR J18293-1213. The dashed straight lines indicate the direction of the rectangle shift due to the influence of interstellar extinction (under the assumption of a standard extinction law; Cardelli et al. 1989) (for more detail, see the text).\label{igrj18293_tsvet}
\end{centering}
\end{figure}

\subsection*{IGR\,J18219-1347}

According to the \emph{UKIDSS} data, the infrared source closest to the localization center of IGR J18219-1347 is the object with coordinates R.A.=18$^h$21$^m$54.766$^s$  Dec=-13\deg47\arcmin26.77\arcsec\ (the distance from the position of the X-ray source is $\simeq0.8$\arcsec, see Fig. 3) with the following infrared magnitudes:  $J=18.00\pm0.05$, $H=16.01\pm0.03$, $K=14.44\pm0.01$ (the standard catalog).

However, a more detailed examination shows that this object is noticeably elongated in the H and K bands. Our studies showed that the PSF is fairly symmetric over the frame Therefore, some elongation of the object clearly indicates that it is either composite in nature or non-pointed. Nevertheless, based on X-ray observations, we can make some assumptions about the nature of IGR J18219-1347. Figure 6 shows the broadband (0.5-100 keV) spectrum of the source obtained from \emph{XRT/Swift} (ObsID. 00031649, 00032285) and \emph{IBIS/INTEGRAL} data. The shape of the spectrum is typical of Galactic binaries and can be fitted by a power law with a slope $\Gamma\simeq 1$ and an exponential cutoff at an energy of $\simeq6$ keV. The absorption $N_H\simeq3\times10^{22}$ measured in the source's spectrum exceeds the matter column density in the Galaxy in this direction. The excess absorption can be produced by matter of the stellar wind from the companion star if it is a giant. Thus, our preliminary conclusion is that this X-ray system is a candidate for high-mass binaries. To ultimately resolve this question, infrared spectroscopic observations of the source should be carried out.

\begin{figure}
\begin{centering}
\includegraphics[width=10cm,bb=15 270 565 690]{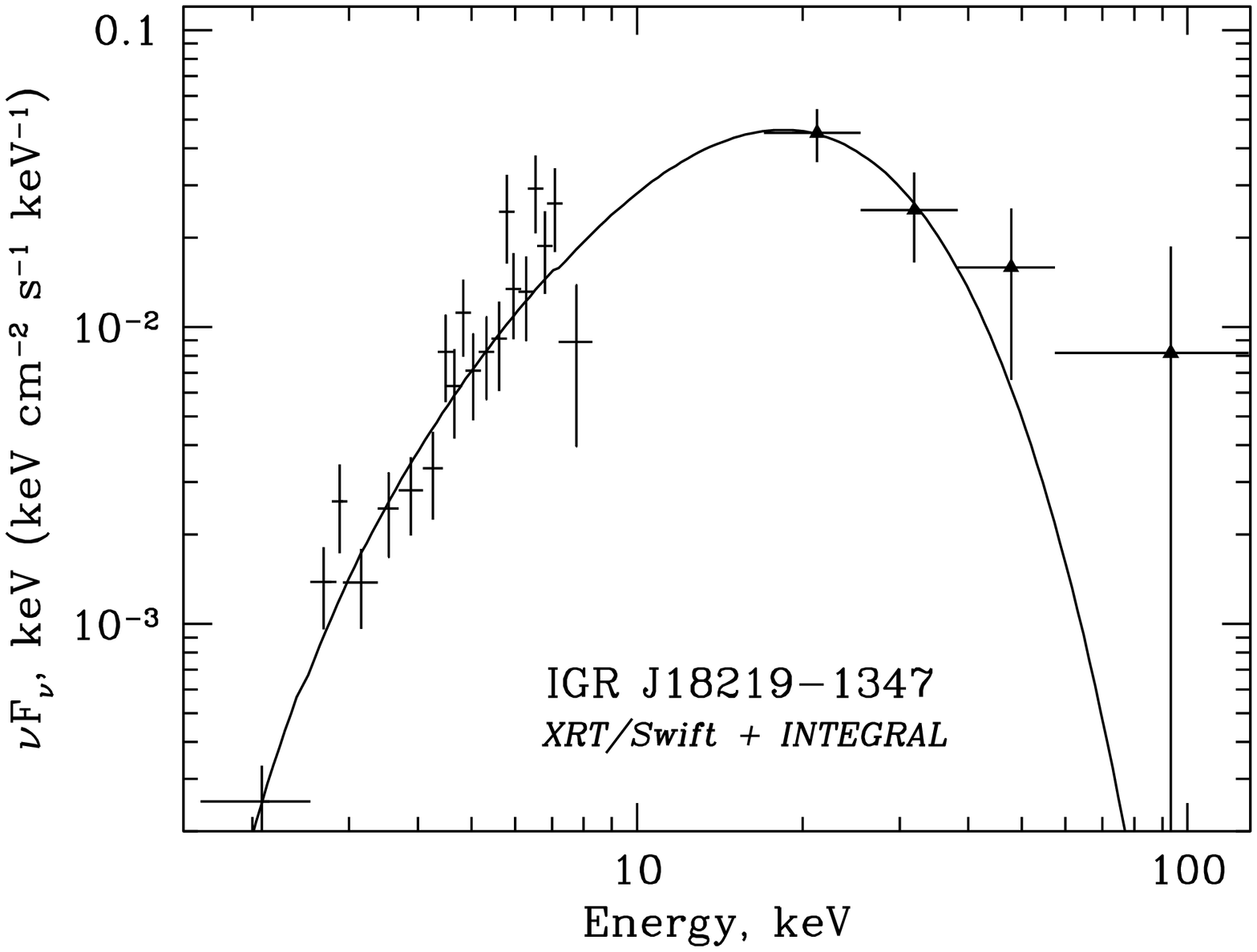}

Fig. 6 Energy spectrum of IGR J18219-1347 from \emph{Swift} (crosses) and \emph{INTEGRAL} (triangles) data. The solid line indicates the best power-law fit with an exponential cutoff and absorption.\label{igrj18219_spectr}
\end{centering}
\end{figure}

\subsection*{IGR\,J17350-2045}

There is a very bright star ($J \approx 9$) in the \emph{2MASS} catalog near the \emph{XMM-Newton} error box of IGR J17350-2045, which makes a proper optical identification of the source from these data impossible. However, using archival data from the SOFI instrument of the \emph{NTT} telescope (the European Southern Observatory, ESO; the observations were performed on May 26, 2007; the exposure time was 4 and 6 s in the H and Ks bands, respectively), we were able to distinguish a faint optical object with coordinates R.A. =17$^h$34$^m$58.834$^s$, Dec=-20$\deg$45\arcmin32.08\arcsec\ (J2000) located at a distance of $\sim$1.8\arcsec\ from the Xray localization center. Our photometric analysis of the data yielded the following magnitudes: $H=14.92\pm0.06$ and $Ks=14.37\pm0.06$ (see. fig.3(d)).
The broadband X-ray spectrum of IGR J17350-2045 obtained from \emph{XMM-Newton} and \emph{INTEGRAL} data in the energy range 0.5-100 keV is shown in Fig. 7. We see from the figure that the data from both observatories are in good agreement with each other and that the spectrum in the entire energy range can be fitted by a simple power law with a slope $\Gamma=1.5\pm0.2$ and absorption $N_H=(18\pm3)\times10^{22}$ cm$^{-2}$. Such a shape of the spectrum and its parameters are typical of active galactic nuclei. Since extragalactic sources efficiently emitting hard X-rays are often also radio sources, we analyzed the radio image of the sky region under study. It turned out that the source NVSS J173459-204533 with a flux of about 13 mJy is present in the \emph{NVSS} sky survey at this location (the difference between the astrometric positions of the radio source and the X-ray source is $\simeq3$\arcsec , which is compatible with their localization errors). Thus, IGR J17350-2045 is most likely an active galactic nucleus. Additional spectroscopic observations are needed to determine its distance.
\begin{figure}
\begin{centering}
\includegraphics[width=10cm,bb=15 270 565 690]{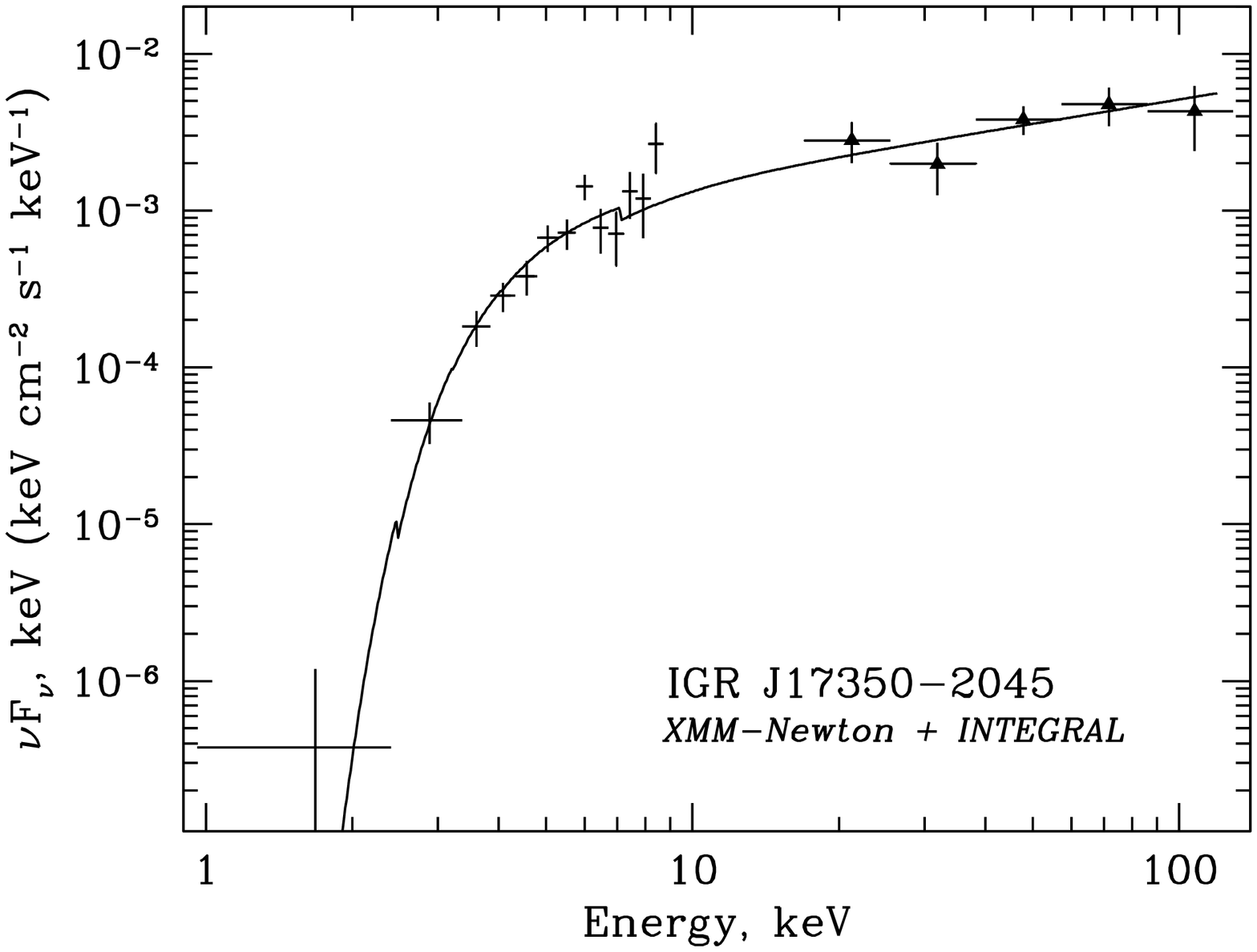}

Fig.7 Energy spectrum of IGR J17350-2045 from \emph{XMM-Newton} (crosses) and \emph{INTEGRAL} (triangles) data. The solid line indicates the best power-law fit with absorption.\label{igrj17350_xmmint}
\end{centering}
\end{figure}

\subsection*{IGR\,J18048-1455}

The hard X-ray source IGR J18048-1455 was discovered in the \emph{IBIS/INTEGRAL} survey of the Galactic plane (Bird et al. 2006; Krivonos et al. 2007). The first attempts to determine the nature of the source were made by Burenin et al. (2006). Based on the detection of a weak $H_\alpha$ emission line, they supposed that the source is a high-mass binary. Subsequently, Masetti et al. (2008) hypothesized that the source was a low-mass X-ray binary. To clarify the nature of the source, we analyzed its XMM-Newton observations on March 3, 2007. The source's spectrum taken in these observations is presented in Fig. 8. We clearly see that there is a set of three emission lines in the spectrum at energies 6-7 keV. Three of them, at 6.7 and 6.9 keV, are ordinary emission lines of highly ionized iron ions in a hot plasma (the plasma temperature determined by fitting the observed spectrum in the range 3-10 keV with the radiation model of a single-temperature plasma is $kT=10\pm1$ keV), while the third line, at $6.38\pm0.02$ keV, is most likely a fluorescent line of neutral (or weakly ionized) iron.

Such a shape of the spectrum allows the nature of the source to be unambiguously determined as an accreting white dwarf. A rather significant photoabsorption with a column density $N_H=(8.1\pm0.7)\times10^{22}$ cm$^{-2}$ is required to properly describe the energy spectrum in the 3-10 keV energy band. This exceeds considerably the column density of the interstellar matter in the same direction ($\sim10^{22}$; Schlegel et al. 1998), implying that the observed absorption in the X-ray band is local in nature. Such a photoabsorption in the binary systems with white dwarfs can be formed in the matter of the accretion stream itself. Such a large photoabsorption typically arises in systems with a high accretion rate and, consequently, with a high X-ray luminosity, $L_{X}\sim 10^{33}-10^{34}$ ergs cm$^{-1}$. From this consideration at the source's observed X-ray flux of $\sim10^{-11}$ erg cm$^{-2}$ s$^{-1}$, we can make an estimation of the distance to the source to be 1-3 kpc. Our analysis of the source's light curve revealed periodic (as far as can be judged from the XMM-Newton observations, with a length of about 13 ks) variability of the signal from the source with a period of $simeq$1440 s (Fig. 9; the probability that the detected signal is random is $\simeq2\times10^{-5}$). The detected variability is most likely associated with the rotation period of the white dwarf and, consequently, IGR J18048-1455 is an intermediate polar.

\begin{figure}
\begin{centering}
\includegraphics[width=10cm,bb=15 270 565 690]{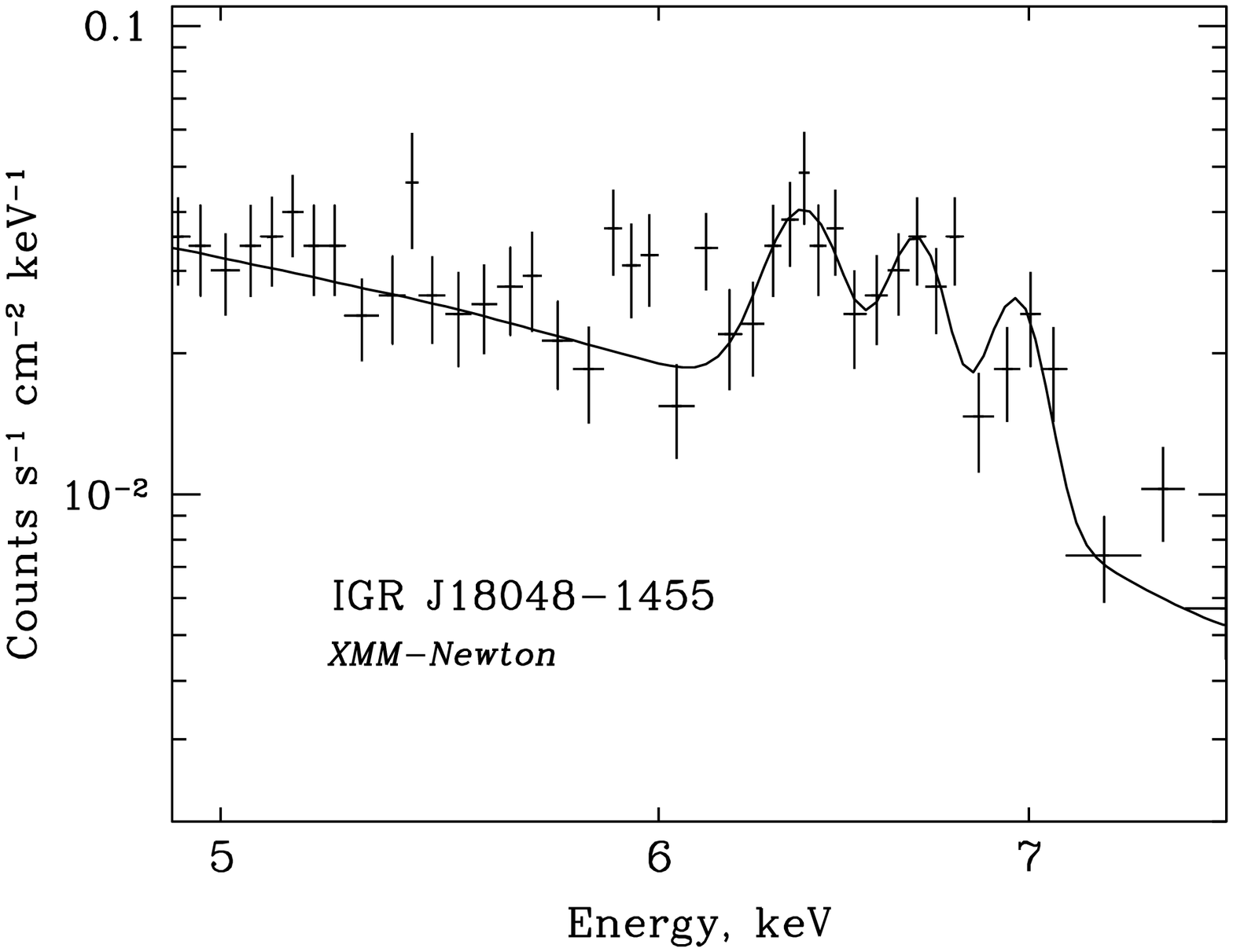}

Fig.8 Portion of the X-ray spectrum for IGR J18048-1455 in the 5-8 keV energy band obtained from \emph{XMM-Newton} data. The triplet of iron lines at energies of 6.4, 6.7, and 6.9 keV is clearly seen. The solid curve indicates the best fit (for more detail, see the text).\label{igrj18048_spec}
\end{centering}
\end{figure}

\begin{figure}
\begin{centering}
\includegraphics[width=11cm,bb=45 185 565 465]{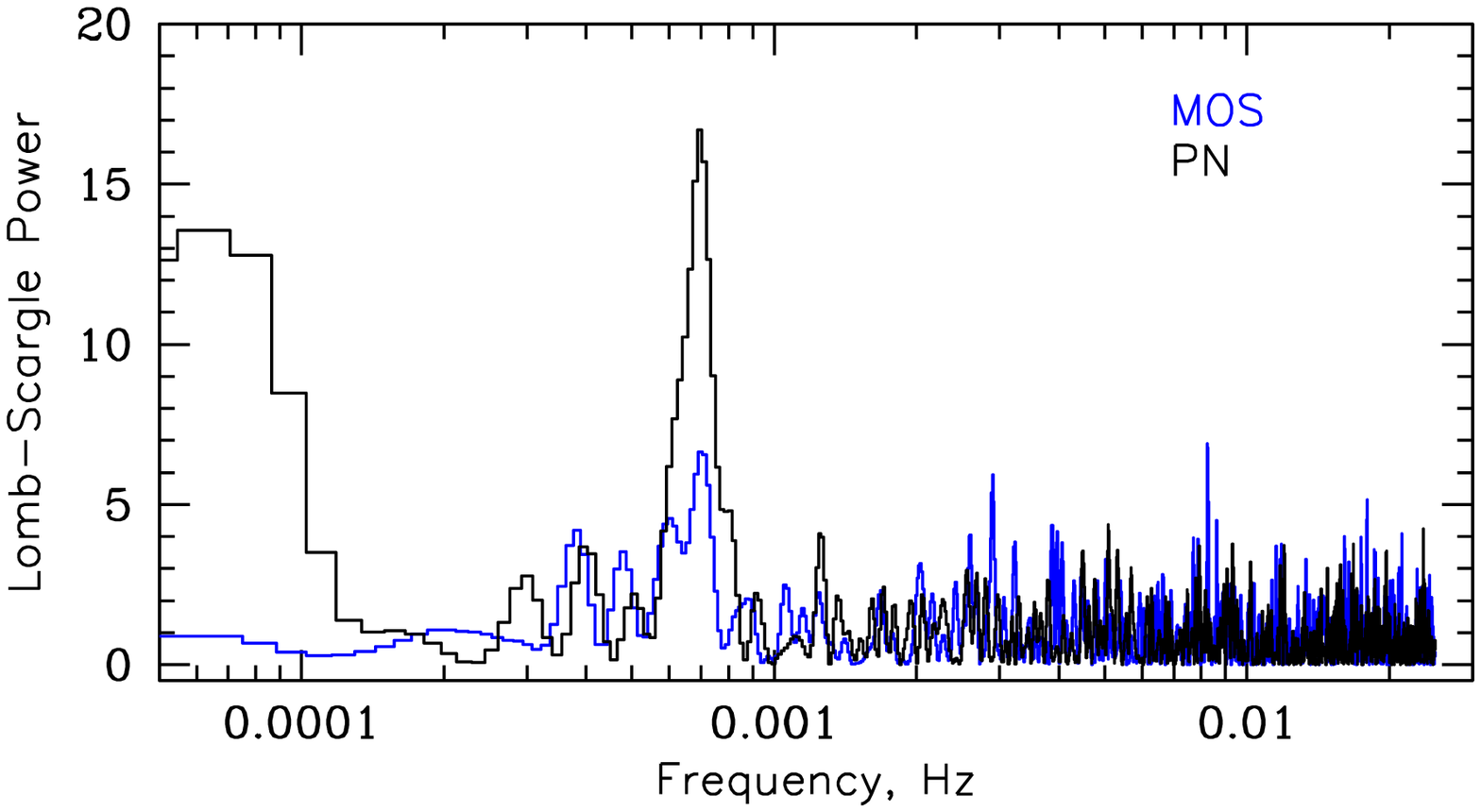}

Fig.9 Lomb-Scargle periodogram for the light curve of IGR J18048-1455 based on data from the $PN$ and $MOS1$ telescopes of the \emph{XMM-Newton} observatory.\label{igrj18048_lcurve}
\end{centering}
\end{figure}

\newpage

\section*{CONCLUSIONS}

We obtained accurate localizations for six hard X-ray sources from the \emph{INTEGRAL} all-sky survey. Using data from the \emph{2MASS} and \emph{UKIDSS} infrared catalogs and from the \emph{NTT} telescope, we have established for the first time the optical counterparts for three of them (IGR J12134-6015, IGR J18293-1213, IGR J17350-2045). The nearest optical object for IGR J18219-1347 probably consists of two sources that cannot be separated with the currently available data. Nevertheless, based on the object's broadband spectrum, we can hypothesize that it belongs to the class of high-mass X-ray binaries. The improvement of the error box for the transient Xray source XTE J1901+014 and the corresponding upper limit on the Ks-band luminosity of the optical star allowed us to confirm the previously made assumptions about a low-mass nature of this system and to derive constraints on its distance and the presumed orbital period. The broadband spectra of IGR J12134-6015 and IGR J17350-2045 obtained from \emph{XMM-Newton}, \emph{Swift}, and \emph{INTEGRAL} data can be well fitted by a simple power law with slopes $\Gamma\simeq0.85$ and $\Gamma\simeq1.5$, respectively. In combination with infrared and radio observations, this suggests an extragalactic nature of both sources; the former is most likely a blazar and the latter is an active galactic nucleus. The discovery of X-ray pulsations from IGR J18048-1455 with a period of $\simeq1440$1440 s and the detection of a triplet of iron lines at energies of 6.4, 6.7, and 6.9 keV in its spectrum allowed the source to be unambiguously classi?ed as a cataclysmic variable, an intermediate polar.

\bigskip

ACKNOWLEDGMENTS

We used data from the archives of the Goddard Space Flight Center (NASA) and the European Southern Observatory. This work was financially supported by the Russian Foundation for Basic Research (project nos. 12-02-01265, 11-02-12285-ofim-2011), the Presidium of the Russian Academy of Sciences (the "Nonstationary Phenomena in the Universe" Program), the Program for Support of Scientific Schools of the President of the Russian Federation (grant no. NSh-5603.2012.2), and the State contract no. 14.740.11.0611.

\section*{REFERENCES}

\begin{enumerate}

\item Bird A., Barlow E., Bazzani L., et al. \apj\, {\bf 636}, 765 (2006)

\item Burenin R., Mesheryakov A., Revnivtsev M., et al. Astronomers Telegram, {\bf 880}, 1 (2006)

\item  Cardelli, J. A., Clayton, G. C., Mathis, J. S., \apj\, {\bf 345}, 245 (1989)

\item Gehrels N., Chinkarini G., Giommi P., et al \apj\, {\bf 611}, 1005 (2004)

\item Iben I.,Jr., Tutukov A.V. \apj, {\bf 284}, 719 (1984)

\item  Karasev D. I.,  Lutovinov A. A., and Grebenev S. A., Astronomy Letters, {\bf 33}, 159, (2007)

\item  Karasev D. I., Lutovinov A. A., and Burenin R. A., Astronomy Letters, {\bf 34}, 753 (2008)

\item  Karasev D. I., Revnivtsev M. G., Lutovinov A. A., and
Burenin R. A., Astronomy Letters, {\bf 36}, 788 (2010)

\item Kasliwal M., Cenko S., Kulkarni S., et al \apj, {\bf 678}, 1127 (2008)

\item  Krimm H., Romano P., Vercellone S., et al Astron. Telegram, {\bf 2375}, 1 (2010)

\item  Krivonos R., Revnivtsev M., Lutovinov A., et al \aap\, {\bf 475}, 775 (2007)

\item  Krivonos R., Tsygankov S., Revnivtsev M., et al \aap\, {\bf 523}, A61, (2010a)

\item R. Krivonos, M. Revnivtsev, S. Tsygankov, et al., Astron. Astrophys. {\bf 519}, A107 (2010b)

\item  Krivonos R., Tsygankov S., Lutovinov A., et al \aap\, {\bf 545}, A27 (2012)

\item  Marshall D., Robin A., Reyle C., et al A\&A, {\bf 453}, 635 (2006)

\item  Masetti N., Mason E., Morelli L., et al  A\&A, {\bf 482}, 113 (2008)

\item  Predehl P., Schmidt J. A\&A, {\bf 293}, 889 (1995)

\item  Remillard R., Smith D., Astron. Telegram, {\bf 88}, 1 (2002)

\item  Revnivtsev M., Postnov K., Kuranov A., Ritter H. \aap\, {\bf 526}, A94 (2011)

\item  Revnivtsev M., Zolotukhin I., Meshcheryakov A., \mnras\, {\bf 421}, 2846 (2012)

\item Schlegel D., Finkbeiner D., Davis M. \apj\, {\bf 525}, 500 (1998)

\item   Smith D., Rampy R., Negueruela I., Astron. Telegram, {\bf 1268}, (2007)

\item  Torrejon J., Negueruela I., Smith D., Harrison T., \aap, {\bf 510}, A61, (2010)

\item van Leeuven F., Fantino E. \aap\, {\bf 439}, 791 (2005)

\item van Paradijs J., McClintock J. \aap\, {\bf 290}, 133 (1994)

\item Wegner W. \mnras, {\bf 270}, 229 (1994)

\item Wegner W. \mnras, {\bf 371}, 185 (2006)

\item Wegner W. \mnras, {\bf 374}, 1549 (2007)

\end{enumerate}


\end{document}